\documentclass[aps,preprint,floatfix,nofootinbib,showpacs]{revtex4-1}
\pdfoutput=1
\usepackage{graphicx,color}
\usepackage{hyperref}
\begin{document}

\title{$WW$ Scattering in the Era of Post Higgs Discovery}
\renewcommand{\thefootnote}{\fnsymbol{footnote}}

\author{
Jung Chang$^1$, Kingman Cheung$^{1,2}$, Chih-Ting Lu$^1$, Tzu-Chiang Yuan$^3$}
\affiliation{
$^1$ Department of Physics, National Tsing Hua University,
Hsinchu 300, Taiwan \\
$^2$ Division of Quantum Phases and Devices, School of Physics, 
Konkuk University, Seoul 143-701, Republic of Korea \\
$^3$ Institute of Physics, Academia Sinica, Nangang, Taipei 11529, Taiwan}
\renewcommand{\thefootnote}{\arabic{footnote}}
\date{\today}

\begin{abstract}
  More evidences have now been collected at the Large Hadron Collider
  suggesting the new 125$\sim$126 GeV boson is likely the long sought
  Higgs boson in the standard model.  One pressing question continued
  being asked by theorists is whether this Higgs boson is a lone
  player responsible for the full electroweak symmetry breaking.
  Current data still allow room for additional Higgs bosons or some
  other UV physics that may play a partial role in electroweak
  symmetry breaking as well.  We use the $WW$ scattering to
  investigate such a possibility, using the two-Higgs-doublet model as
  a prototype. The $WW$ scattering becomes strong when the extra Higgs
  bosons are very heavy.  We study the sensitivity of the {\it
    partially strong} $WW$ scattering signals at the 13 TeV Large
  Hadron Collider.
\end{abstract}
\pacs{14.80.Bn, 14.80.Cp, 12.60.Fr, 12.15.Ji}
\maketitle

\section{ Introduction}  
A new particle with mass of 125$\sim$126 GeV was discovered 
at the Large Hadron Collider (LHC) in July 2012 \cite{atlas,cms}. 
This may be the long sought Higgs boson of the standard model (SM), 
which was proposed in 1960s
\cite{higgs}, or one of the Higgs bosons beyond the SM. For example,
supersymmetric theories, little-Higgs models, and other extended Higgs
sector such as the two-Higgs-doublet models (2HDM) all contain a multitude
of neutral as well as charged Higgs bosons. The current data still
contain large uncertainties that these various extensions of the  
SM cannot be confirmed or ruled out decisively.  
Based on the data on the signal strengths of
all decay channels of the Higgs boson, it is therefore important to constrain
various couplings of the Higgs boson.  Indeed, several precision studies of
the Higgs boson appeared recently, 
either in model-independent approach \cite{CLT}
or in specific models (e.g., two-Higgs-doublet models \cite{2hdm1}).

One of the most useful constraints from the global fitting of the Higgs
boson couplings is the one to a pair of $W/Z$ bosons. The current
data constrain  \cite{CLT} 
\begin{equation}
\label{Cv}
C_v \equiv \frac{ g_{hWW}}{g_{hWW}^{\rm SM}} = 0.96 \,^{+0.13}_{-0.15} \; \; .
\end{equation}
The central value is close to $1$, which means that the observed Higgs boson
leaves only little room for the existence of 
another Higgs boson or some unknown UV
physics responsible for the electroweak symmetry breaking (EWSB).  
If $C_v$ is exactly equal to $1$, it
means that the observed Higgs boson will completely account for the EWSB.
We do not need another Higgs boson, or if another Higgs boson exists it 
has nothing to do with the EWSB, for example in the inert Higgs doublet model. 
Nevertheless, it is not unreasonable that the value of $C_v$ could 
deviate from the central value by $-2\sigma$,
then the $C_v$ could be as low as $0.66$. One certainly needs more data
to reduce the error.

If the $hWW$ coupling is less than its SM value, there must be something
heavier, could be as heavy as a few TeV, to complete the EWSB. 
The simplest realization of this scenario is the 2HDM, in which
the light CP-even Higgs boson $h$ is at 
125-126 GeV while the heavy CP-even Higgs boson $H$ is at 1-2 TeV. 
These two CP-even Higgs bosons couple to the vector boson with reduced 
strengths 
$g_{hWW} =\sin (\beta - \alpha) g_{hWW}^{\rm SM}$ and 
$g_{HWW} =\cos(\beta - \alpha) g_{hWW}^{\rm SM}$ such that 
$g_{hWW}^2 + g_{HWW}^2 = (g_{hWW}^{\rm SM})^2$,
where $\tan\beta$ is the ratio of the VEVs of the two doublets and 
$\alpha$ the mixing angle
between  the two CP-even neutral Higgs bosons.
At low energy only one light CP-even Higgs boson is relevant. 
One can then parametrize all the UV effects with all the heavier Higgs 
bosons being integrated out 
by an effective Lagrangian as presented in Ref.\cite{Giudice}.

As is well known, scattering of the longitudinal components of the weak
gauge bosons is a useful probe of the EWSB sector \cite{WW,equiv}.
The scattering amplitudes 
with purely gauge contributions grow with energy as
$s/m_W^2$, where $s$ is the squared center-of-mass (CM) energy of the
$W_LW_L$ system.  In the SM with a light Higgs boson, the amplitude
will be completely unitarized by the Higgs boson.  Once $\sqrt s$ goes
beyond the light Higgs boson mass, the scattering amplitude will no
longer grow like $s/m_W^2$.  
On the other hand, if the $hWW$ coupling deviates from the SM value,
even by a small amount, the terms growing like $s/m_W^2$ in the
scattering amplitude would become strong after hitting the light Higgs
pole.  Furthermore, 
if the scale of the UV part is far enough from the light Higgs
boson, the onset of strong $W_LW_L$ scattering between the light Higgs
mass and the UV scale should be discernible at the LHC, as was
demonstrated for a generic extended Higgs sector in \cite{CCY} as well
as for an extra hidden $Z'$ model in \cite{CCHY}.
This temporal growth of $W_LW_L$ scattering amplitudes in the
immediate range of energy is of immense interests to the LHC experiments,
in particular with its upgrade to 13-14 TeV.
Previously, the calculation was done using the naive effective $W$
approximation \cite{ewa}.  In this work, we extend our previous work
\cite{CCY} to include the full calculations with detector simulations.

In the full calculation of $q q \to q q W^+ W^-$, there are (i)
vector-boson fusion (VBF) diagrams and (ii) non-VBF diagrams, e.g., 
the $W$ bosons simply radiate off the external quark legs. 
The non-VBF diagrams do not involve the dynamics in the EWSB sector, 
and thus should be suppressed by devising appropriate kinematical cuts. 
The $WW$ fusion can be extracted by the presence of two energetic 
forward jets. We can impose selection cuts to select jets in 
forward rapidity and high energy region \cite{kin1}. Furthermore, if
we demand the leptonic decay of the vector bosons, there will be very
few jet activities in the central rapidity region \cite{kin2}. Previous
studies in the context of strongly interacting EWSB sector were performed in
Ref.~\cite{madison}. Similar selection cuts can be applied here for
partially strong $W_LW_L$ scattering.  

The organization of this paper is as follows. In Sec. II, we first
briefly review some details how the bad energy behavior in the
$W_LW_L$ scattering amplitudes in SM is completely cancelled among
the gauge and Higgs diagrams. We then discuss how the modified gauge-Higgs
coupling may lead to incomplete cancellation and thus the partial
growth in the scattering amplitudes in the intermediate energy range.
Using the 2HDM as an illustration we present our numerical results to
support this partially strong $W_LW_L$ scattering for the 13 TeV
LHC in Sec. III.  We conclude in Sec. IV.

Note that the use of the 2HDM is only for simplicity and renormalizablitiy. 
The main point here
is that the model could account 
for the light CP-even Higgs boson observed at the
LHC. This Higgs boson is partially responsible for the EWSB and the other 
part of the EWSB is rather heavy. 
The 2HDM has at least six independent free parameters and 
certainly has enough freedom to allow us to implement this scenario.  
We are looking at the window between this light Higgs boson and 
the heavy UV part where 
the $W_L W_L$ scattering may reveal the nature of the EWSB sector at the LHC.

A previous work on using $WW$ scattering to investigate the anomalous
$g_{hWW}$ coupling can be found in Ref.~\cite{hj}.
Another interesting approach is to determine the relative longitudinal to
transverse production of the vector bosons by measuring the polarization of 
the vector bosons \cite{tao}.

\section{$WW$ Scattering Amplitudes}

Let us begin by recalling the derivation of a covariant form for the
longitudinal polarization 4-vector $\epsilon_L^\mu (p)$ of the vector
boson, say the $W$ boson.  The leading piece is directly proportional
to $p^\mu/m_W$.  We can write it as
\begin{equation}
\epsilon^\mu_L(p) = p^\mu / m_W + v^\mu(p)
\end{equation}
with 
\begin{equation}
v^\mu(p) \simeq - \frac{m_W}{2 {p^0}^2} ( p^0, - \vec{p}) \sim O(m_W/E_W) \; .
\end{equation}
Since this form of $v^\mu$ is not covariant, so the calculation
involving $v^\mu$ would be cumbersome. Nevertheless, in the
center-of-mass frame of the incoming $W^+(p_1) W^- (p_2)$ pair, where
$\vec{p}_1 = - \vec{p}_2$, we can express
\begin{equation}
v^\mu (p_1) = - \frac{2 m_W}{s} p_2^\mu 
\end{equation}
and so the polarization 4-vector $\epsilon_L^\mu(p_1)$ can be expressed as 
\begin{equation}
\label{4vector}
\epsilon_L^\mu (p_1) = \frac{p_1^\mu}{m_W} - \frac{2 m_W}{s} p_2^\mu 
\end{equation}
and similarly
\begin{equation}
\epsilon_L^\mu (p_2) = \frac{p_2^\mu}{m_W} - \frac{2 m_W}{s} p_1^\mu 
\end{equation}
where $s = (p_1 + p_2)^2$. For outgoing $W^+(k_1) W^- (k_2)$ pair,
simply make the substitution of $(p_1,p_2) \to (k_1,k_2)$ to obtain
the covariant form for their polarization vectors.

Next consider the process 
$W^+ (p_1) W^- (p_2) \to W^+ (k_1) W^- (k_2) $, which has contributing
Feynman diagrams of $\gamma,Z$ in both $s$ and $t$ channels and a 
4-point vertex, as well as the Higgs boson exchange in $s$ and $t$ channels.
The amplitudes for the gauge diagrams are given by
\begin{eqnarray}
\label{gaugeamp}
i {\cal M}_t^{\gamma+Z} &=& -i g^2\left( 
    \frac{s^2_W}{t} + \frac{c_W^2}{t - m_Z^2} \right )
   \left[ 
    (p_1+k_1)^\mu  \epsilon(p_1) \cdot \epsilon(k_1)
 -   2 k_1 \cdot \epsilon(p_1) \; \epsilon^\mu(k_1) 
 -   2 p_1 \cdot \epsilon(k_1) \; \epsilon^\mu(p_1) \right ] \nonumber \\
 &&\times \left[
    (p_2+k_2)_\mu  \epsilon(p_2) \cdot \epsilon(k_2)
 -   2 k_2 \cdot \epsilon(p_2) \; \epsilon_\mu(k_2) 
 -   2 p_2 \cdot \epsilon(k_2) \; \epsilon_\mu(p_2) \right ] \nonumber  \;\; , 
\\
i {\cal M}_s^{\gamma+Z} &=& -i g^2 \left(
    \frac{s^2_W}{s} + \frac{c_W^2}{s - m_Z^2} \right )
   \left[ 
    (p_1-p_2)^\mu  \epsilon(p_1) \cdot \epsilon(p_2)
 +   2 p_2 \cdot \epsilon(p_1) \; \epsilon^\mu(p_2) 
 -   2 p_1 \cdot \epsilon(p_2) \; \epsilon^\mu(p_1) \right ] \nonumber \\
 &&\times \left[
    (k_2-k_1)_\mu  \epsilon(k_1) \cdot \epsilon(k_2)
 -   2 k_2 \cdot \epsilon(k_1) \; \epsilon_\mu(k_2) 
 +   2 k_1 \cdot \epsilon(k_2) \; \epsilon_\mu(k_1) \right ] \nonumber  \;\; ,\\
i {\cal M}_4 &=& i g^2 \left[ 
  2 \epsilon(p_2) \cdot \epsilon(k_1)\; \epsilon(p_1) \cdot \epsilon(k_2)
  - \epsilon(p_2) \cdot \epsilon(p_1)\; \epsilon(k_1) \cdot \epsilon(k_2)
  - \epsilon(p_2) \cdot \epsilon(k_2)\; \epsilon(p_1) \cdot \epsilon(k_1)
   \right ]
         \nonumber \; \; .
\end{eqnarray}
Substituting the form of the longitudinal polarization vectors into
the above amplitudes, the leading term of order $O(E^4/m_W^4)$ of each
amplitude is
\begin{eqnarray}
\label{gaugeamp2}
i {\cal M}_t^{\gamma+Z} &=& -i \frac{g^2}{4 m_W^4} \left[ (s-u)t - 3 m_W^2(s-u)
  + \frac{8 m_W^2}{s} u^2 \right ]  \nonumber  \; \; , \\
i {\cal M}_s^{\gamma+Z} &=& -i \frac{g^2}{4 m_W^4} \left[ s(t-u) - 3 m_W^2(t-u)
   \right ] \nonumber  \; \; , \\
i {\cal M}_4 &=& i \frac{g^2}{4m_W^4} \left[ s^2 + 4 st + t^2 - 4 m_W^2(s+t)
 - \frac{8 m_W^2}{s} u t \right ] \nonumber \; \; .
\end{eqnarray}
The gauge structure ensures that the cancellation of $O(E^4/m_W^4)$
terms \footnote{In an extra hidden $Z'$ model, it has been shown in
  \cite{CCHY} that even the $O(E^4/m_W^4)$ terms are not cancelled and
  may lead to partially strong $W_LW_L$ scattering as well.}.  The sum
of the gauge diagrams are left with terms proportional to
$O(E^2/m_W^2)$:
\[
 i {\cal M}^{\rm gauge} = i {\cal M}_t^{\gamma+Z} + i {\cal M}_s^{\gamma+Z}  + 
i {\cal M}_4 = 
 -i \frac{g^2}{4 m_W^2} u +O( (E/m_W)^0 ) \; \; .
\]
Suppose the $hWW$ coupling is merely a fraction $C_v$ of its SM value
as defined in Eq.(\ref{Cv}).
The contributions from the Higgs diagrams are
\begin{eqnarray}
\label{amp-Higgs}
i {\cal M}^{\rm Higgs} &=& - i \frac{ C_v^2 g^2}{4 m_W^2} \left[
  \frac{ (s - 2 m_W^2)^2}{s -m_h^2} + \frac{(t - 2 m_W^2)^2}{t -m_h^2} 
  \right ] \; , \nonumber \\
 & \simeq &
  i \frac{C_v^2 g^2}{4 m_W^2} \, u  \;,
\end{eqnarray}
in the limit of $s \gg m_h^2, m_W^2$.  Only if $C_v$ is exactly equal
to $1$ as in SM can the bad energy-growing term be delicately
cancelled between the gauge diagrams and the Higgs diagrams.
Historically an upper bound of the SM Higgs mass of $m_h^2 < 4 \pi
\sqrt 2 / G_F$ was first deduced based on the unitarity constraint on
the $W_LW_L$ scattering \cite{WW}.  Nowadays more useful theoretical
lower and upper bounds for the Higgs mass $129 < m_h {\rm (GeV)} <
180$ can be obtained by considering the vacuum stability
\cite{vacuumstability} and perturbativity \cite{pertubativity} of the
SM scalar potential.
Nevertheless, back to our own track.  In some extended models that the
light Higgs boson has only a fraction of the SM coupling strength
(i.e. $C_v < 1$), one expects the total scattering amplitude to keep
growing with $s$ after hitting the light Higgs pole at 125-126 GeV.
We expect the UV part of the EWSB sector will come in eventually to
unitarize the amplitude at sufficiently high energy to restore
unitarity. It was shown that the violation of unitarity occurs at
$\sqrt{s_{WW}} = 1.2, 1.7, 2.7, 3.8 $ TeV for a modified $hWW$
coupling with $C_v = 0, 0.71, 0.89, 0.95$, respectively \cite{CCY}.

As alluded already in the introduction, the simplest realization 
of the above scenario of temporal growth of $W_LW_L$ scattering 
amplitude is the 2HDM, in which $C_v $ is given by $\sin (\beta - \alpha )$.
The heavier neutral Higgs boson
couples to the weak gauge boson with a reduced strength $g_{HWW} = \cos
(\beta-\alpha) g_{HWW}^{\rm SM}$ such that it can unitarize the rest
of the growing amplitudes when $s_{WW} > m_H^2$.  
We will use this scenario in 2HDM to investigate the sensitivity at the
LHC.

\section{Experimental Cuts for VBF and Numerical Results}

The central issue for the experimental detection of $WW$ scattering is to
separate the VBF diagrams among all the other non-VBF ones.  In the
VBF diagrams, each of the initial quarks radiates a $W/Z$ boson, which
further scatters into the final state $W/Z$ bosons.  The unique feature
of this process is that the scattered quark is very energetic, carrying
almost all the energy of the incoming quark and very forward
\cite{kin1,kin2}. Furthermore, if we demand the leptonic decays of the $W$
and $Z$ bosons, there will be very little hadronic activities in the
central rapidity region.  Therefore, the signature includes (i) the
appearance of two energetic forward jets with large spatial
separation, and (ii) the leptonic decay products of the $W$ or $Z$
bosons are enhanced at the large invariant mass region.

Based on these features we impose the following experimental cuts for
the two jets in selecting the VBF events:
\begin{equation}
\label{jcut1}
E_{T_{j1,j2}} > 30 \; {\rm GeV},\;\;  |\eta_{j1,j2}| < 4.7 ,\;\; 
\Delta \eta_{12} = \vert \eta_{j1} - \eta_{j2} \vert > 3.5, \;\; 
\eta_{j1} \eta_{j2} < 0 \;\; ,
\end{equation}
where $E_{T_{j1,j2}}$ and $\eta_{j1,j2}$ are the transverse energies and 
pseudo-rapidities respectively of the two forward jets, and 
\begin{equation}
\label{jcut2}
M_{jj} >  500 \;{\rm GeV} 
\end{equation}
on their invariant mass $M_{jj}$ at $\sqrt{s}=13$ TeV.
This set of cuts is more or less the same as those used by
CMS~\cite{cms-fp} and ATLAS \cite{atlas-fp} in their searches for fermiophobic 
Higgs boson using VBF.
The cuts for the leptonic decay modes 
$W\to \ell \nu_\ell$ and $Z\to \ell^+ \ell^-$ for
each of the $W^+ W^-$, $W^\pm W^\pm$, $W^\pm Z$, and $ZZ$ channels are
slightly different, which we list separately in Table~\ref{cuts}.
We sum over the charged leptons $\ell=e,\mu$.

\begin{table}[tb!]
\caption{\small \label{cuts}
Leptonic cuts on the leptonic decay products of the diboson channels:
$W^+ W^-$, $W^\pm W^\pm$, $W^\pm Z$, and $ZZ$.}
\centering
\begin{ruledtabular}
\begin{tabular}{llll}
$ W^+ W^-$ & $ W^\pm W^\pm$ & $ W^\pm Z$ & $ Z Z$ \\
\hline
$p_{T_{\ell}} > 100$ GeV &  $p_{T_{\ell}} > 100$ GeV & 
$p_{T_{\ell}} > 100$ GeV & $p_{T_{\ell}} > 50$ GeV  \\
$\left|y_\ell \right| < 2$ & $\left|y_\ell \right| < 2$ & 
$\left|y_\ell \right| < 2$ & $\left|y_\ell \right| < 2$ \\
$M_{\ell^+\ell^-} > 250$ GeV &  $M_{\ell^\pm\ell^\pm} > 250$ GeV & 
$M_{3\ell} > 375$ GeV & $M_{4\ell} > 500$ GeV 
\end{tabular}
\end{ruledtabular}
\end{table}

We set the mass of the heavier CP-even Higgs boson to be 2 TeV, which
basically at the margin of the LHC reach. The charged and the CP-odd 
Higgs bosons are not relevant to the $WW$ scattering here.
Therefore, the only relevant parameter to this study is $\sin(\beta-\alpha)$,
which we shall use $0.5,0.7,0.9$ as illustrations.

We use {\sf MADGRAPH} 5 \cite{madgraph5} to perform the full parton-level
calculations, including the decays of the $W$ and $Z$ bosons. 
Then we turn on {\sf PYTHIA} 8.1 \cite{pythia81} for parton showering and 
jet radiation, and use {\sf PYTHIA-PGS} \cite{pythia-pgs} to 
perform detector simulation to provide jet and lepton reconstruction.

\begin{figure}[t!]
\centering
\includegraphics[width=6.8in]{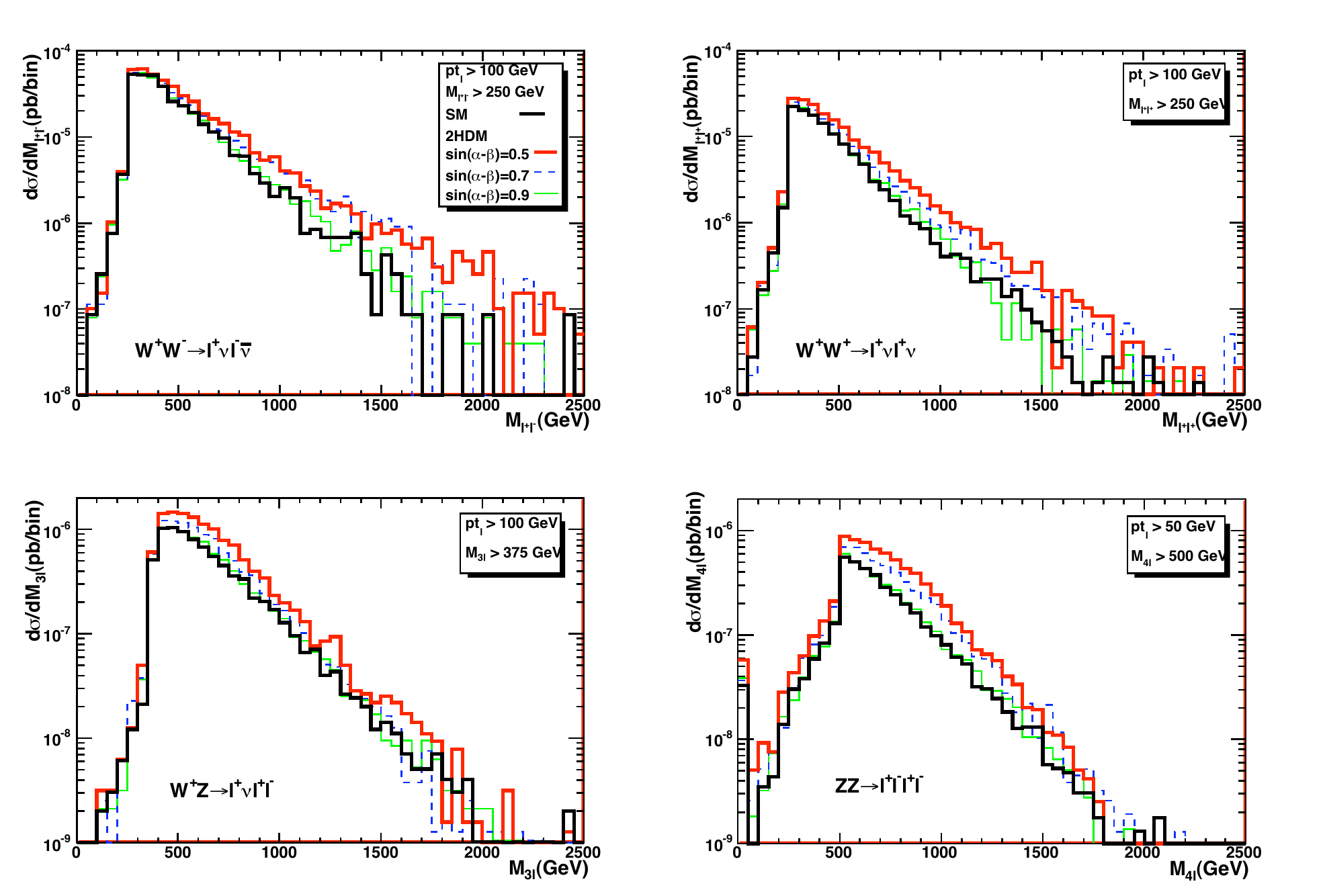}
\caption{\small \label{fig1}
Invariant-mass distributions of the charged leptons in
various diboson channels $W^+ W^-, W^+ W^+, W^+ Z, ZZ$ for 
$\sin(\beta-\alpha)= 0.5,0.7,0.9$.}
\end{figure}

We expect that the enhancement in the differential cross section in the large
invariant-mass region of the vector-boson pair will be manifested at the
large invariant-mass of its decay products, e.g., $M_{\ell\ell}$ in both
$W^\pm W^\pm$ and $W^+ W^-$ channels, and $M_{3\ell}$ and $M_{4\ell}$ 
in $WZ$ and $ZZ$ channels, respectively (see Table~\ref{cuts}).
We show the invariant-mass distributions of the charged leptons in
various diboson channels
$W^+ W^-, W^+ W^+, W^+ Z, ZZ$ for $\sin(\beta-\alpha)= 0.5,0.7,0.9$ 
as well as the SM
in Fig.~\ref{fig1}. In the figures, we show only $W^+ W^+$ and $W^+ Z$ channels
since $W^- W^-$ and $W^-Z$ are relatively smaller. 
In Table~\ref{x-sec}, we show the cross sections for all channels 
after the leptonic and jet cuts in various diboson channels for 
$\sin(\beta-\alpha) = 0.5,0.7, 0.9$ and the SM.

\begin{table}[tbh!]
\caption{\small \label{x-sec}
Cross sections in fb in various diboson channels under the jet cuts 
in Eqs.~(\ref{jcut1}) and (\ref{jcut2}), and leptonic cuts listed in
Table~\ref{cuts}. 
}
\centering
\begin{ruledtabular}
\begin{tabular}{lllll}
           & \multicolumn{4}{c}{Cross Sections (fb) } \\
Channels  & $\sin(\beta-\alpha)= 0.5$ & $0.7$ & $0.9$ &  SM ($C_v=1$) \\
\hline
$W^+W^- \to \ell^+\nu\ell^-\bar\nu$ & $0.51$ & $0.46$ & $0.40$ & $0.39$ \\
$W^+ W^+ \to \ell^+\nu\ell^+\nu$ & $0.20$ & $0.17$ & $0.14$ & $0.14$ \\
$W^- W^- \to \ell^-\bar\nu\ell^-\bar\nu$ & $0.083$ & $0.075$ & $0.070$ & $0.069$
  \\
$W^+ Z \to \ell^+\nu\ell^+\ell^-$& $0.016$ & $0.013$ & $0.011$ & $0.010$ \\
$W^- Z \to \ell^-\bar\nu\ell^+\ell^-$& $1.0\times 10^{-2}$ & 
   $8.5\times 10^{-3}$ & $7.6\times 10^{-3}$ & $7.4\times 10^{-3}$ 
\\
$ZZ \to \ell^+\ell^-\ell^+\ell^-$ & $8.4\times 10^{-3}$ & $6.4\times 10^{-3}$ 
             & $4.6\times 10^{-3}$ & $4.4\times 10^{-3}$ 
\end{tabular}
\end{ruledtabular}
\end{table}

The difference between the cross section of the SM and the one with
$\sin(\beta-\alpha) \neq 1$ is the signal of enhancement due to the
deviation in the $g_{hWW}$ coupling.  The largest enhancement happens
in the $W^+ W^-$ and $ZZ$ channels. 
In the $W^+ W^-$ channel, the
enhancement is $(0.51-0.39)/0.39 \approx 0.31$ and $(0.46-0.39)/0.39
\approx 0.18$ for $\sin(\beta-\alpha)=0.5$ and $0.7$, respectively;
while in the $ZZ$ channel the enhancement is $(8.4-4.4)\times 10^{-3}
/ 4.4\times 10^{-3} = 0.91$ and $(6.4-4.4)\times 10^{-3} / 4.4\times
10^{-3} = 0.45$ for $\sin(\beta-\alpha)=0.5$ and $0.7$, respectively.
Because of the overall smallness of the $ZZ$ channel compared with
other channels, even if we can collect the planned 300 fb$^{-1}$
luminosity at the LHC, the event rate for $ZZ \to 4 \ell$ is too small
for detection. On the other hand, the event rate for $W^+ W^- \to
2\ell 2 \nu$ is sufficient for detection at the LHC.

\section{Conclusion}

In this work, we have demonstrated that detailed studies of longitudinal
weak gauge boson scattering at the LHC can provide useful hints of new
physics at a higher scale, despite only a light Higgs boson has been
discovered at the LHC.  If unitarity is only partially fulfilled by
the light Higgs boson, the $WW$ scattering cross sections must be growing as
energy increases before it reaches the other heavier Higgs bosons or
other UV completion to achieve the full unitarization. This partial
and temporary growth of the cross sections can be palpable at the LHC
provided that the UV part resides at a sufficiently high scale.  
On the other hand, if the UV part is within the reach of the LHC 
energies, the $WW$ scattering can also reveal it as 
a bump in the invariant mass distribution.
This can be realized in a number of multi-Higgs-doublet models, e.g, 2HDM.
Our approach of using longitudinal weak gauge boson scattering is more
direct and perhaps more efficient to probe the EWSB.  Partial growth
in the $WW$ scattering cross sections can be a generic feature in many
extensions of the SM.  Detection of such a behavior at the LHC would
be fascinating.  Perhaps Higgs is not a lone player.

{\it Acknowledgments.}  This work was supported in part by the
National Science Council of Taiwan under two grants
99-2112-M-007-005-MY3 and 101-2112-M-001-005-MY3 as well as
the WCU program through the KOSEF funded by the MEST
(R31-2008-000-10057-0).


\end{document}